\documentclass[aps,prx,reprint,floatfix,superscriptaddress,footinbib,twocolumn]{revtex4}
\usepackage{xcolor}
\usepackage{microtype}
\usepackage[bookmarks=true,
bookmarksnumbered=false, 
bookmarksopen=false, 
colorlinks=true,
linkcolor=blue]{hyperref}
\definecolor{webblue}{rgb}{0, 0, 0.5} 
\hypersetup{colorlinks=true,urlcolor=blue,citecolor=blue}
\usepackage{graphicx}
\usepackage{subfigure}
\usepackage{url}
\usepackage{hyperref}
\usepackage{inconsolata}
\usepackage{amsmath}
\usepackage[capitalize]{cleveref}
\usepackage{braket}
\usepackage{outlines}
\usepackage{enumitem}
\usepackage{soul}
\usepackage{color}

\usepackage{bm} 

\usepackage[utf8]{inputenc}
\usepackage{siunitx,booktabs}
\usepackage{multirow}
\usepackage{amssymb}

\usepackage{relsize}

\usepackage{soul}
\usepackage{tabularx}

\begin{document}

\begin{abstract}
We demonstrate the existence of novel interaction effects in hole-doped semiconductor quantum wells which are connected to dramatic changes in the Fermi surface geometry occurring upon variation of the doping. We present band structure calculations showing that quantum wells formed in $p$-type cubic semiconductors develop perfectly nested Fermi surfaces at a critical hole density $p \sim 1/d^2$ set by the width $d$ of the quantum well. Nesting gives rise to competing superconducting and charge or spin density wave order, which we analyze using the perturbative renormalization group method. The correlated phases may be created or destroyed by tuning the hole density towards or away from the critical density. Our results establish $p$-type semiconductor quantum wells as a platform for novel correlated phases, which may be precisely controlled using electrostatic gating and external magnetic fields.
\end{abstract}

\title{Nested Fermi surfaces and correlated electronic phases in hole-doped semiconductor quantum wells}

\author{Tommy Li}
\affiliation{Dahlem Center for Complex Quantum Systems and Fachbereich Physik, Freie Universit\"{a}t Berlin, Arnimallee 14, 14195 Berlin, Germany}
\author{Julian Ingham}
\affiliation{Physics Department, Boston University, Commonwealth Avenue, Boston, MA 02215, USA}
\author{Harley D. Scammell}
\affiliation{School of Physics, University of New South Wales, Sydney 2052, Australia}
\affiliation{Australian Research Council Centre of Excellence in Future Low-Energy Electronics Technologies, University of New South Wales, Sydney 2052, Australia}

\date{\today}
\maketitle

\section{Introduction}

Two-dimensional (2D) hole-doped semiconductor systems, which host spin-$\frac{3}{2}$ charge carriers, are known for having an extremely strong, electrically controllable spin-orbit interaction which makes them ideal candidates for the study of novel transport effects and a platform for spintronics. While the effect of spin-orbit coupling on single-particle physics in 2D semiconductor quantum wells has been studied in several recent experimental and theoretical works \cite{Yeoh2014,Nichele2015,Li2016,Marcellina2017,Marcellina2018,Bladwell2019,Cullen2021}, the evolution of the Fermi surface geometry with hole density and its consequences for many-body effects have remained unexplored. The recent observation of coexisting density wave and superconductivity in transition metal dichalcogenides \cite{Barrera2018,Lu2018,Lu2015,Yang2018,Ye2012,Song2021}, kagome metals \cite{Ortiz2020,Zhu2021,Chen2021,Ortiz2021,Ni2021,Chenb2021,Liang2021}, and superconductivity with proximate correlated insulators in spin-orbit coupled Moir\'{e} materials \cite{Wang2020,Jin2021,Xu2020,Huang2021,Miao2021,Ghiotto2021,Shi2021} have stimulated an emergence of interest in the effect of spin-orbit coupling on correlated phases, yet the possibility of many-body instabilities originating from the spin-orbit coupling in 2D spin-$\frac{3}{2}$ systems has not been considered. Hole-doped semiconductor quantum wells therefore currently provide an intriguing opportunity for investigation.

In this work we demonstrate the existence of correlated electronic states which arise due to the spin-$\frac{3}{2}$ nature of the positively charged carriers characteristic of hole-doped semiconductors. We perform band structure calculations that show that, at certain values of the carrier density, the 2D Fermi surfaces exhibit perfect nesting over a range of materials, due to a transition in which the curvature of the Fermi surface changes sign at high symmetry points in $k$-space. Accounting for the presence of an arbitrarily weak Coulomb interaction, we find that competing superconducting and charge (CDW) or spin (SDW) density wave instabilities emerge, with either SDW or coexisting CDW and SDW orders present when the Fermi surface is spin degenerate, and a pure CDW when the spin degeneracy is lifted by either broken time reversal or spatial inversion symmetry. The nested regions do not cover the entire Fermi surface, leading to a Fermi surface reconstruction in which the nested portions are gapped and the unnested portions form Fermi arcs, so that the ordered phases are metallic. Nesting remains possible in an in-plane magnetic field directed along a high symmetry axis, and the nesting densities exhibit a dependence on the magnetic field, allowing the identification and study of the correlated ground state via magnetotransport; in this case only CDW order is present. In diamond-structure semiconductors in zero magnetic field,  SDW order always emerges upon tuning of the hole density close to the nesting density; thus electrical gates provide a highly sensitive control over the magnetic ordering, a property which could prove useful for spintronic applications \cite{Jungwirth2016}.

\section{Hole band structure}

We consider a semiconductor quantum well, which consists of a semiconducting layer typically of width $d \approx 20$ nm, sandwiched between two insulators, with charge carriers confined to a 2D plane. Hole-doped heterostructures formed in cubic semiconductors are described by the Luttinger Hamiltonian \cite{Luttinger1955,Dresselhaus1956,Winkler2003}
\begin{gather}
H = \frac{p^2}{2m}+ T^{\mu\nu\alpha\beta}p_\mu p_\nu S_\alpha S_\beta +  V(z)
+ \nonumber \\
C_1^{\mu\nu\alpha\beta} S_\mu p_\nu p_\alpha p_\beta + C^{\mu\nu\alpha\beta}_2 p_\mu S_\nu S_\alpha S_\beta \ \ , 
\label{LuttingerHamiltonian}
\end{gather}
where $S_\mu$ are the spin-$\frac{3}{2}$ operators, $T,C_1,C_2$ are fourth-rank tensors originating from spin-orbit coupling, and $V(z)$ is a confining potential. Taking a coordinate system aligned with the cubic axes, the tensor $T$ has two independent nonvanishing components, $T^{xxxx}= T^{yyyy} = T^{zzzz}<0$ and $T^{xyxy} = T^{yzyz} = T^{zxzx}<0$. The final two terms in (\ref{LuttingerHamiltonian}) are the Dresselhaus interactions, which appear only in materials with broken bulk inversion symmetry. The tensors $C = \{C_2, C_3\}$ only have one independent nonvanishing component $C^{xxyy} = C^{yyzz} = C^{zzxx} = - C^{xxzz} = -C^{yyxx} = -C^{zzyy}$.

\begin{figure}[t]
\includegraphics[width = 0.5\textwidth]{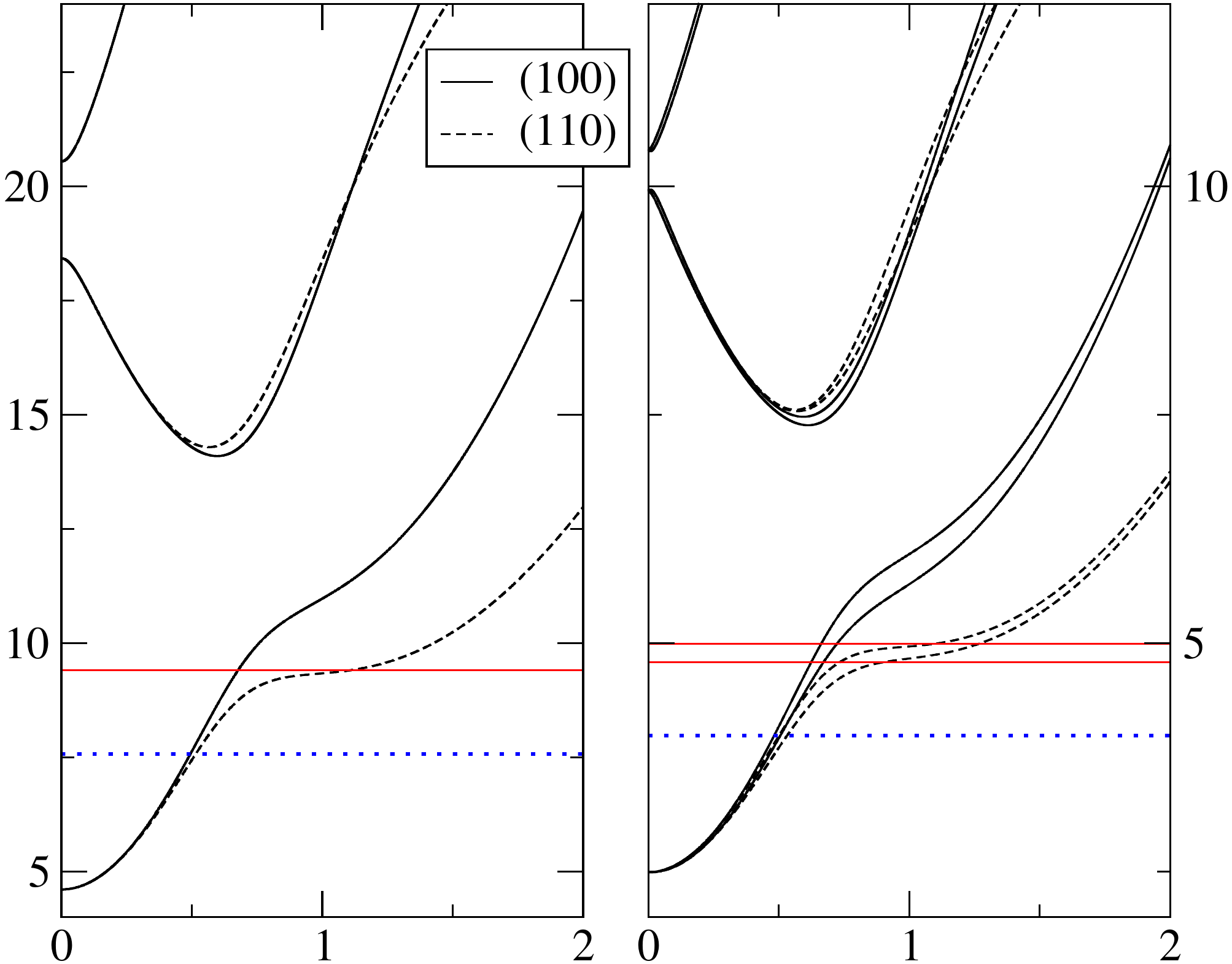}
\caption{Dispersion of 2D subbands, $\varepsilon_n(|\bm{k}|)$ in Ge (left) and GaAs (right) semiconductor quantum wells of width 20 nm along the (100) and (110) directions (solid and dashed lines respectively). The horizontal blue dashed lines show the values of the Fermi energy for typical a 2D hole density $p = 10^{11}$ cm$^{-2}$, while the horizontal solid red lines show the values of the Fermi energy at the critical density at which the curvature of the Fermi surface changes sign, which in Ge (left) is $p \approx 2.56\times 10^{11}$ cm$^{-2}$ and in GaAs (right) are $p \approx 1.99\times 10^{11}$/$2.92\times 10^{11}$ cm$^{-2}$ for the larger/smaller spin split Fermi surfaces.  Energy and momentum are plotted in units of meV and $\pi/d$ respectively.}
\label{fig:disp_Ge_GaAs}
\end{figure}

We consider the case where a confining potential quantizes motion in the $z\parallel (001)$ direction, leaving the charge carriers free to move in the [001] plane. The single particle eigenstates possess definite 2D momentum $\bm{k} = (k_x,k_y)$ and form a discrete series of bands corresponding to the quantization of the transverse modes. We perform band structure calculations for the case of an infinite square well of width $d = 20$ nm. In Fig. ~\ref{fig:disp_Ge_GaAs} we plot the 2D dispersion of Ge (left) and GaAs (right) quantum wells, $\varepsilon_n(\bm{k})$, $n = 1, 2, \dots$ as a function of $|\bm{k}|$ along the (100) (solid lines) and (110) (dashed lines) directions. We observe that, for low densities, the dispersion of the lowest bands is parabolic and isotropic, however as the momentum becomes of the order $k \sim\pi/d$, the dispersion becomes both highly anisotropic and nonparabolic due to the level repulsion between the lowest two pairs of 2D bands. The anisotropy of the Fermi surface is due to the tensor $T^{\mu\nu\alpha\beta}$ appearing in the Luttinger Hamiltonian (\ref{LuttingerHamiltonian}) lacking spherical symmetry,  $T^{xxxx} \neq T^{xyxy}$. The horizontal lines indicate the Fermi energy for various hole densities $p$. The dashed blue lines correspond to a typical doping $p = 10^{11}$ cm$^{-2}$; the solid red line in the left panel corresponds to $p = 2.56\times 10^{11}$ cm$^{-2}$, and the solid red lines in the right panel correspond to $p = 1.99\times 10^{11}$ and $2.92\times 10^{11}$ cm$^{-2}$.

The level repulsion and anisotropy of the 2D dispersion in Ge and GaAs quantum wells may be understood straightforwardly from the structure of the Luttinger Hamiltonian (\ref{LuttingerHamiltonian}). The spin-independent terms provide an isotropic contribution to the dispersion that is quadratic in energy, thus only the spin-dependent terms contribute to the effect. We note that, at small in-plane momenta, the bands in GaAs are nearly twofold degenerate, which shows that the Dresselhaus interaction is negligible for $k_x=k_y=0$, thus the splitting of the 2D bands for small in-plane momentum is entirely due to the term $T^{\mu\nu\alpha\beta}p_\mu p_\nu S_\alpha S_\beta \rightarrow T^{zzzz}p_z^2 S_z^2$ in (\ref{LuttingerHamiltonian}), which for an infinite square well results in an energy splitting $\Delta \varepsilon= 2T^{zzzz} \pi^2/d^2$ between the lowest two pairs of subbands at $k_x=k_y=0$. The bands have definite $S_z^2$, with the lowest and second lowest pair possessing maximum ($S_z^2=\frac{9}{4}$) and minimum ($S_z^2 =\frac{1}{4}$) out-of-plane polarization along the growth axis respectively, since $T^{zzzz}<0$ \cite{Winkler2003}. As the in-plane momentum is increased, terms mixing states with different $S_z^2$, i.e. those containing $S_x,S_y$, become significant. The lowest pair of bands disperses upwards, while the second lowest disperses downwards, leading to an anticrossing at $|\bm{k}|\approx k_z= \pi/d$ which is the wavevector related to transverse quantum confinement.

\begin{figure}[t]
\includegraphics[width = 0.45\textwidth]{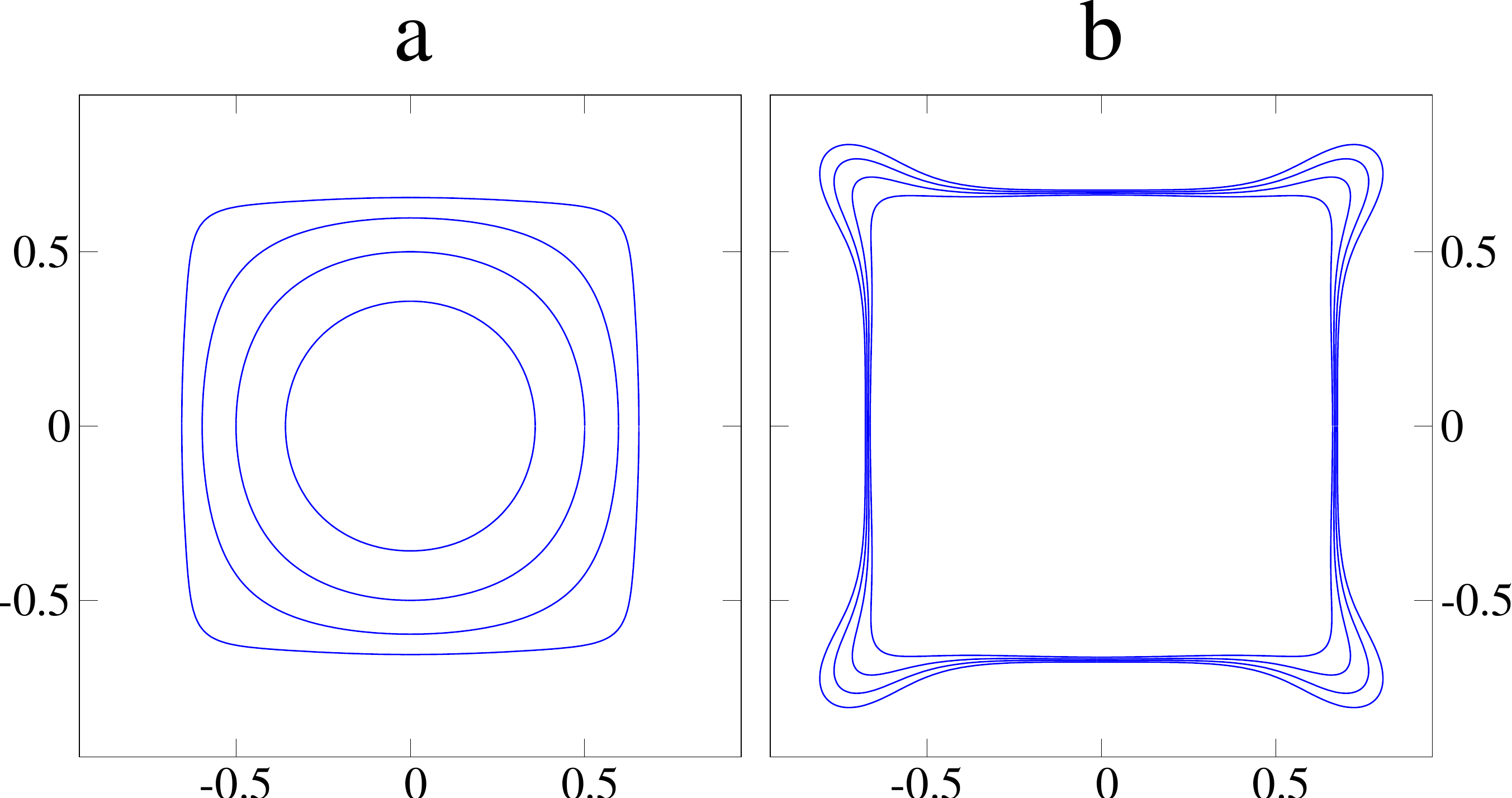}
\caption{Evolution of the Fermi surface of a Ge quantum well with increasing hole doping, with hole densities (in units of $10^{11}$ cm$^{-2}$) (\emph{left}):
$p= 5.13, 1.03, 1.54, 2.05$, (\emph{right}): $p = 2.18, 2.31, 2.43, 2.56$. These densities correspond to $p = 0.2p^*,0.4p^*, 0.6p^*,0.8p^*$ for the left and $p^* = 0.85p^*, 0.9p^*, 0.95p^*, p^*$ where $p^*$ is the doping at which the curvature of the sides of the Fermi surface changes sign (as defined in the text).}
\label{fig:FS_Ge}
\end{figure}

At the anticrossing, the level repulsion between the two lowest pairs of bands is considerably stronger in the (110) direction than in the (100) direction (shown in the solid and dashed lines respectively in Fig ~\ref{fig:disp_Ge_GaAs}). As the Fermi energy is increased to the anticrossing, the Fermi surface becomes progressively more distorted due to the anisotropy of the dispersion. In Fig. ~\ref{fig:FS_Ge} we plot the hole Fermi surfaces in a 2D Ge quantum well at various hole densities. We observe the evolution of the Fermi surface from almost perfectly circular at low densities to progressively more square. As the hole density approaches a critical value $p^* = 2.56 \times 10^{11}$ cm$^{-2}$, the curvature of the sides of the Fermi surface $v_\perp^{-1}\partial^2 \varepsilon/\partial k_\parallel^2$ changes sign (with $v_\perp, k_\parallel$ being the velocity and momenta transverse and longitudinal to the Fermi surface) and the sides of the Fermi surface become perfectly flat. We observe that almost perfect nesting of the Fermi surface occurs over a range of densities.

\begin{figure}[t]
\includegraphics[width = 0.45\textwidth]{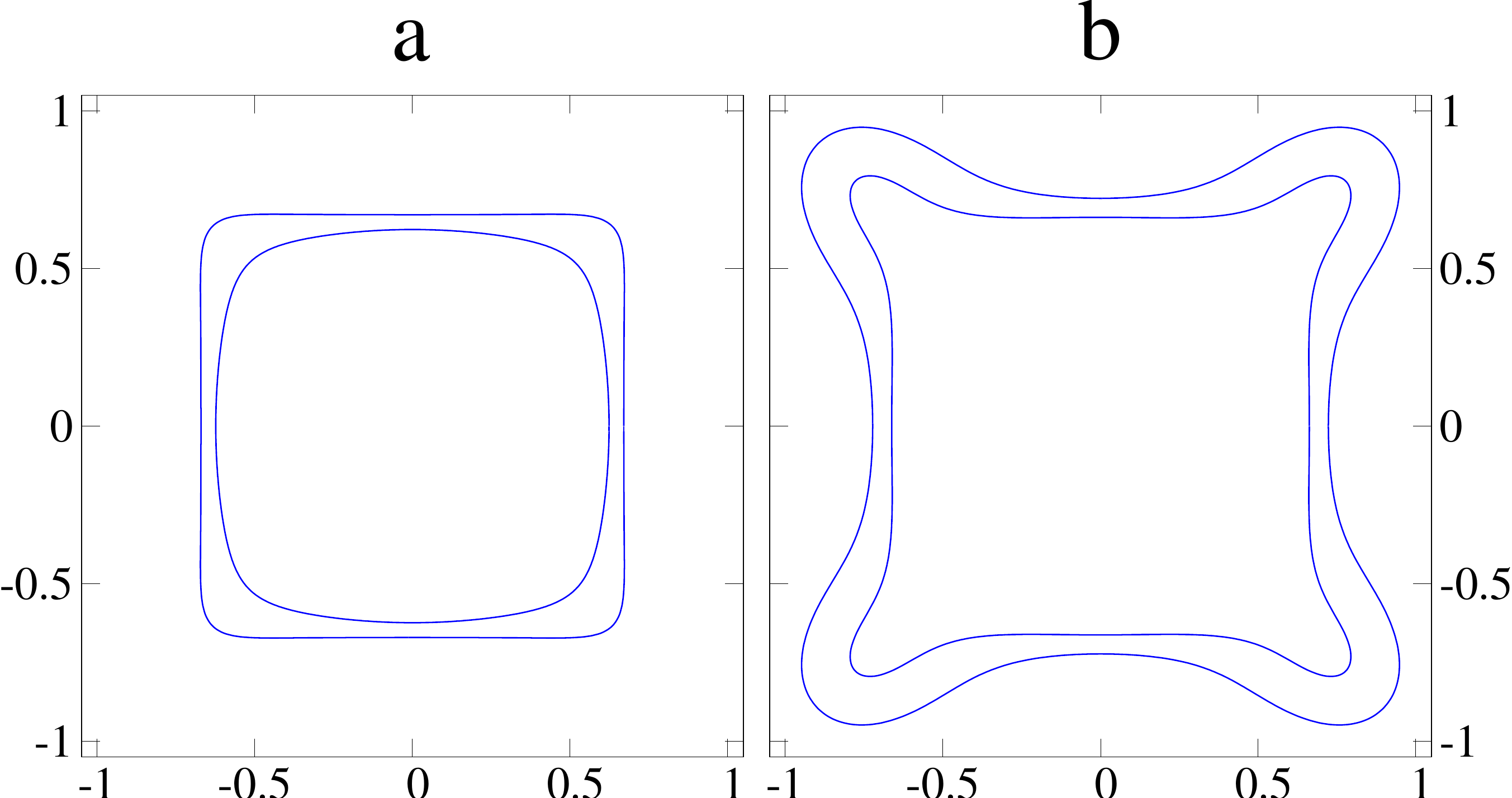}
\caption{The spin-split Fermi surfaces in a GaAs quantum well in zero magnetic field, with hole densities (a): $p = 1.99\times 10^{11}$ cm$^{-2}$, at which the larger Fermi surface is nested, and (b) $p \approx 2.91\times 10^{11}$ cm$^{-2}$, at which the smaller Fermi surface is nested.}
\label{fig:FS_GaAs}
\end{figure}

In Fig. \ref{fig:FS_GaAs} we show the Fermi surfaces in a GaAs quantum well at the hole densities (a) $p = 1.99 \times 10^{11}$ cm$^{-2}$, (b) $p=2.91 \times 10^{11}$ cm$^{-2}$. Due to bulk inversion asymmetry, the bands are no longer spin degenerate, but the same transition between convex and concave Fermi surfaces occurs for each band at different densities.

The nesting of the Fermi surfaces in both Ge and GaAs systems reflects shared features of the 2D dispersion arising from the spin structure of the Luttinger Hamiltonian \ref{LuttingerHamiltonian} common to hole-doped cubic semiconductors. At small in-plane momenta, the bands are approximately isotropic and parabolic, giving rise to circular energy contours. The quantum states in the lowest band have approximately definite $S_z^2 = \frac{9}{4}$. As the momentum is sufficiently increased, the terms in the Hamiltonian containing the in-plane momenta $k_x,k_y$ compete with the 2D quantization (corresponding to wavevector $k_z = \pi/d$) which splits the energy of the $S_z^2=\frac{9}{4}$ and $S_z^2 = \frac{1}{4}$ states, and both the level repulsion and cubic anisotropy become important and eventually distort the energy contours into a four-lobed shape (Fig. ~\ref{fig:FS_Ge}b). For some density between the circular and lobed shape, the energy contours become square, implying there is always a transitional hole density $p \sim 1/d^2$ with exact nesting. The terms in the Hamiltonian (\ref{LuttingerHamiltonian}) responsible for this effect are quadratic in the in-plane spin components, $S_x^2, S_y^2, \{S_x,S_y\}$, which are proportional to the identity matrix for spin-$\frac{1}{2}$ systems but nontrivial for spin-$\frac{3}{2}$. Thus we may directly connect the nesting phenomenon to the combination of spin-$\frac{3}{2}$ nature of holes and 2D confinement. We should expect this to occur generally among quantum wells formed in hole-doped cubic semiconductors, and we indeed find nested Fermi surfaces for Si, InAs, InSb, CdTe and ZnSe, for which we provide figures (Figs. ~\ref{sfig:Si}--\ref{sfig:ZnSe}) in the Supplementary Material. In all cases, assuming a well width of $d = 20$ nm, nesting occurs within an order of magnitude of $p\sim 1/d^2 =  2.5\times 10^{11}$ cm$^{-2}$. The robustness of the existence of nesting densities may be connected to the change in sign of the curvature of the sides of the Fermi surface $v_\perp^{-1}\partial^2 \varepsilon/\partial k_\parallel^2$, which we show in Fig. ~\ref{sfig:concavity} in the Supplementary Material.

\begin{figure}[t]
\includegraphics[width=0.45\textwidth]{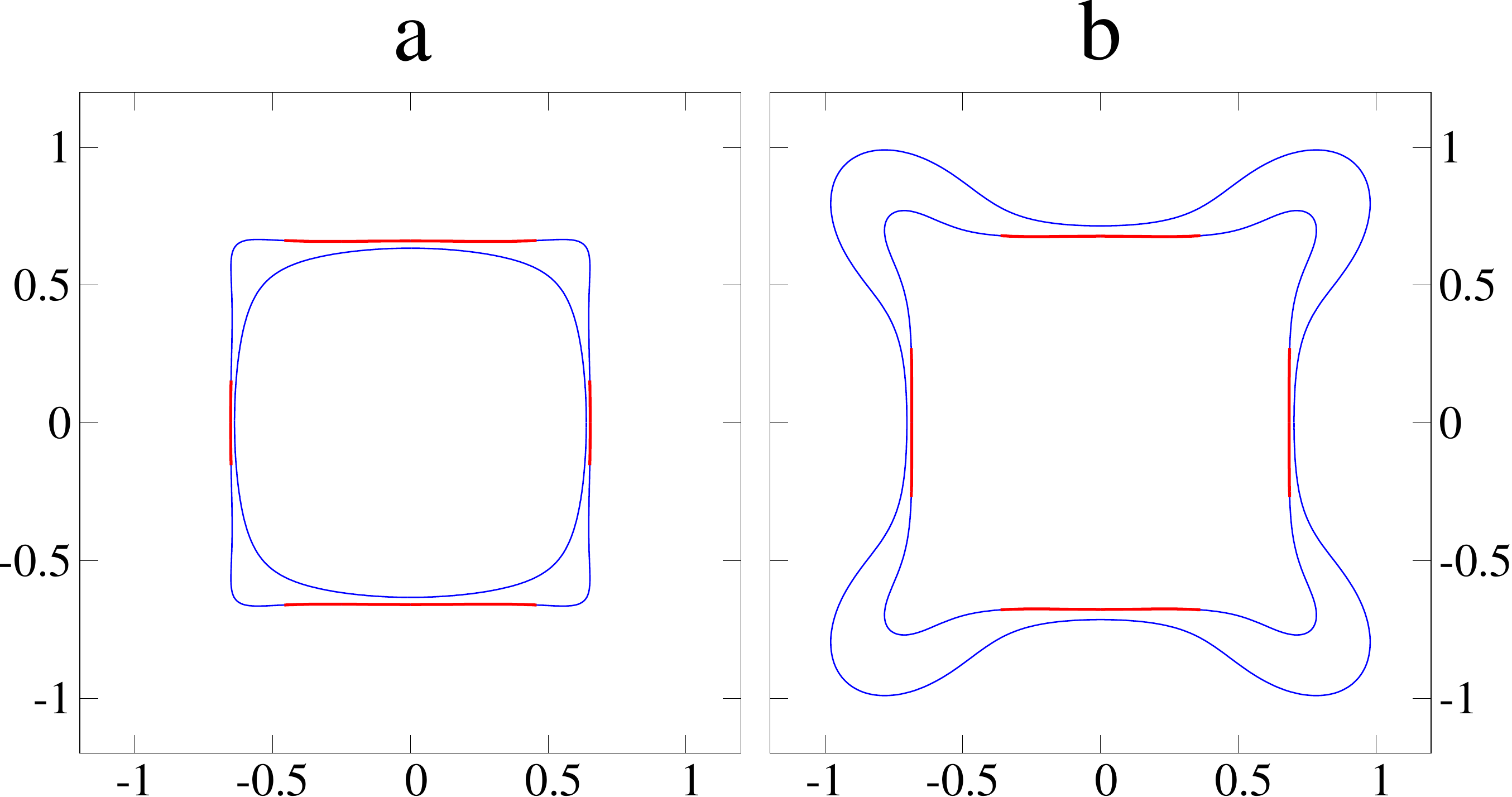}
\caption{Nested Fermi surfaces for a 20 nm Ge quantum well in a 1T magnetic field applied along the (010) direction, at hole densities (a) $1.97\times 10^{11}$ cm$^{-2}$, (b) $3.00 \times 10^{11}$ cm$^{-2}$. Portions of the Fermi surface on which the normal momentum coordinate varies by less than 0.25\% are highlighted in red.}
\label{fig:FS_Ge_1T}
\end{figure}

Due to the cubic symmetry of the bulk Hamiltonian, the Fermi surface is symmetric under reflections about the (100) and (010) axes as well as fourfold rotations, for confinement in the [001] plane. Thus, all four sides are simultaneously nested at the critical densities. We may also consider the situation in an in-plane magnetic field applied along a high symmetry axis, which breaks the fourfold symmetry of the Fermi surface. For diamond semiconductors, the Fermi surface possesses two mirror axes parallel and perpendicular to the magnetic field, however fourfold rotation symmetry is broken, and the twofold spin degeneracy is lifted. In Fig. ~\ref{fig:FS_Ge_1T} we show nested Fermi surfaces of a Ge quantum well in a 1T magnetic field along the (010) direction at two densities, corresponding to a situation when the larger and smaller Fermi surfaces are nested. We find that, for the densities chosen, the sides of the Fermi surface perpendicular to the magnetic field exhibit nesting over a greater region than the sides parallel to the field. However, we also found that at slightly different densities, the opposite scenario was also possible.

\begin{figure}[b]
\includegraphics[width = 0.45\textwidth]{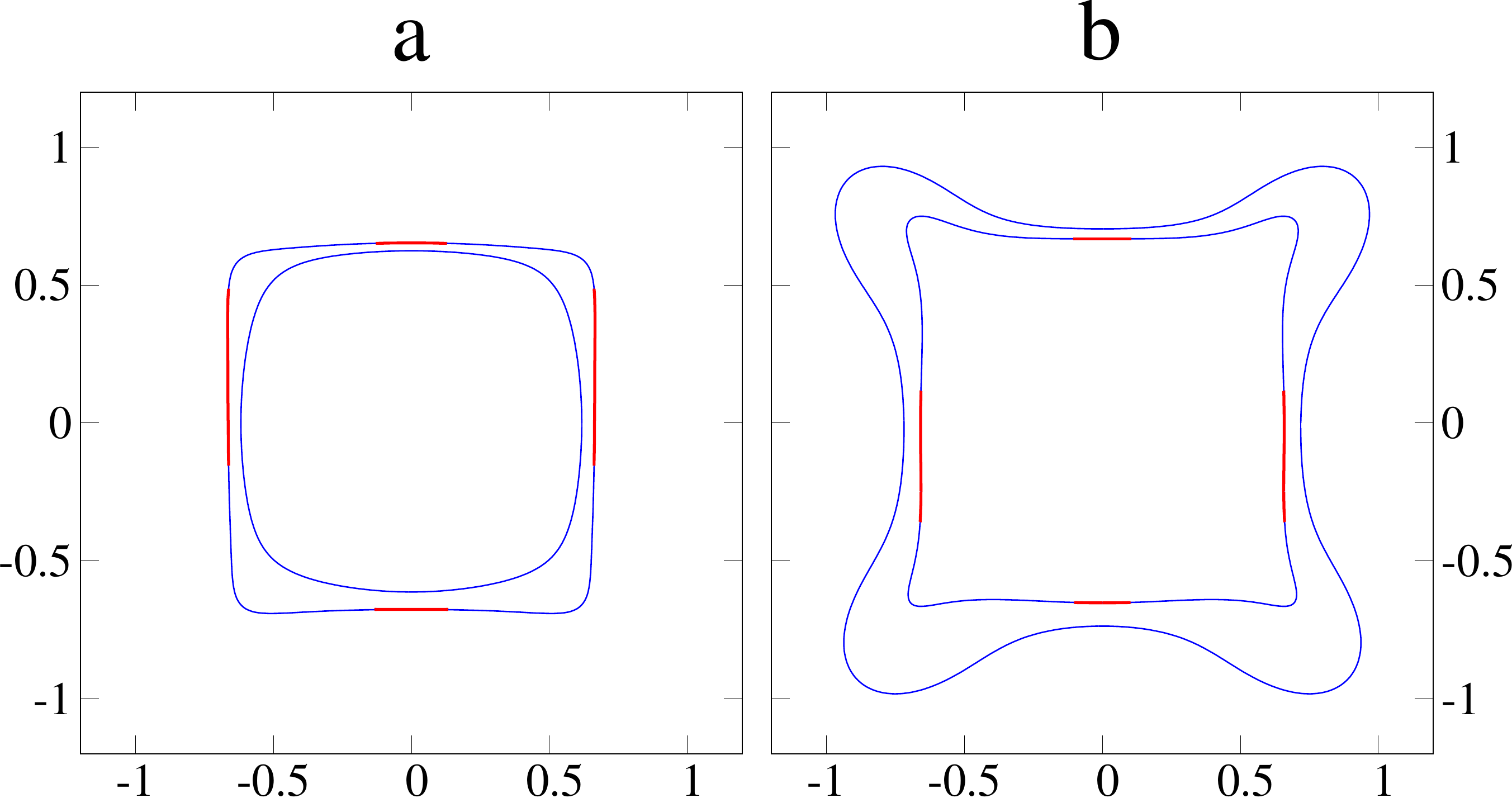}
\caption{Nested Fermi surfaces for a 20 nm GaAs quantum well in a 1T magnetic field applied along the (010) direction, at densitie, (a) $p = 1.94\times 10^{11}$ cm$^{-2}$, (b) $2.83\times 10^{11}$ cm$^{-2}$. Portions of the Fermi surface on which the normal momentum coordinate varies by less than 0.25\% are highlighted in red.}
\label{fig:FS_GaAs_1T}
\end{figure}

In zincblende semiconductors, bulk inversion asymmetry results in a lack of mirror symmetry about the axis transverse to the magnetic field, however mirror symmetry about the direction parallel to the field remains. Among the five zincblende semiconductors we studied (GaAs, InAs, InSb, CdTe, ZnSe), all except InSb exhibited transitions at which the curvature of at least one pair of opposite sides of the Fermi surface changed sign in a 1T magnetic field. In the case of CdTe and ZnSe, significant nesting occurred for all four sides of the Fermi surface, however in the cases of GaAs and InAs, densities were found for which significant nesting occurred only for one pair of sides. In Fig. ~\ref{fig:FS_GaAs_1T} we show the nested Fermi surface of a GaAs quantum well in a 1T magnetic field applied along the (010) direction, at densities for which nesting is exhibited in the larger/smaller Fermi surfaces in the left/right panels. Portions of the Fermi surface on which the normal momentum coordinate varies by less than 0.25\% are highlighted in red. We show the nested Fermi surfaces in a 1T magnetic field for the remaining zincblende semiconductors in the Supplementary Material.

So far we have presented results for the case where the confining potential $V(z)$ is an infinite square well. We have observed qualitatively identical behavior for generic quantum wells which are symmetric along the growth axis, however nesting can be significantly reduced by the asymmetry of the quantum well in zincblende structures, even at zero magnetic field. We present a comparison of the effect of structural inversion asymmetry in Ge and GaAs quantum wells in Fig. ~\ref{sfig:Rashba} in the Supplementary Material.

\section{RG analysis}

\begin{figure}[t]
\includegraphics[width =0.45\textwidth]{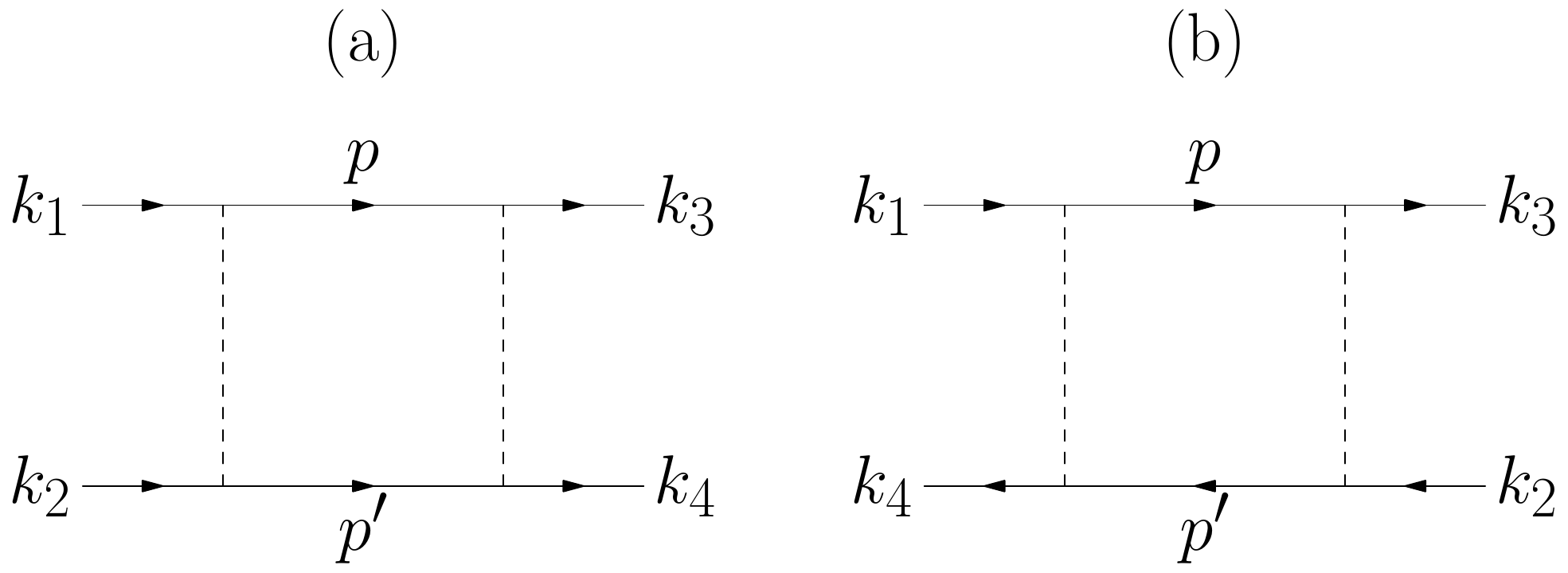}
\caption{One-loop corrections to the interaction vertex.}
\label{fig:diagrams}
\end{figure}

The existence of nested Fermi surfaces suggests a dramatic enhancement in the role of interactions, as is well known from the theory of the fermionic renormalization group \cite{Shankar1994}. We define an effective theory by restricting the allowed momenta $\bm k$ to lie within a cut-off of width $\Lambda$ about the two Fermi surfaces (which are degenerate when both inversion and time-reversal symmetry are present but otherwise split), and study instabilities of the Fermi liquid phase by analyzing the RG flow of the effective couplings $\Gamma^\Lambda_{\sigma_1\sigma_2\sigma_3\sigma_4}(\bm{k}_1,\bm{k}_2,\bm{k}_3,\bm{k}_4)$ describing scattering between two holes with initial momenta $\bm{k}_1,\bm{k}_2$ and final momenta $\bm{k}_3,\bm{k}_4$, and in the initial/final Fermi surfaces indexed by $\sigma_1,\sigma_2$/$\sigma_3,\sigma_4$. We calculate the RG flow of the effective couplings by identifying logarithmic divergences in perturbative corrections to the scattering vertex; many-body instabilities are signaled by a divergence in the effective couplings at low energies ($\Lambda\rightarrow 0$). We begin by presenting a general argument why, irrespective of the specific interacting model, nesting in our systems always leads to an instability, due to the divergence of certain scattering amplitudes at low energies, and later introduce a toy model which allows us to explore the phase diagram.

Noting that the bare values of interactions involving large momentum transfer are suppressed due to the $1/q$ momentum dependence of the Coulomb interaction as well as the overlap between spin--$\frac{3}{2}$ wavefunctions in the initial and final hole states, we might expect that these interactions grow slower under the RG flow and thus the strongest instabilities are associated with small momentum transfer processes with $\sigma_1=\sigma_3,\sigma_2=\sigma_4$. To second order in the interactions, the RG flow is driven by corrections to the scattering vertex associated with the virtual processes shown in Fig. ~\ref{fig:diagrams}, which involve intermediate states consisting of two holes with momenta $\bm{p},\bm{p}'$ (Fig. ~\ref{fig:diagrams}a) or a hole and an electron with momenta $\bm{p},-\bm{p}'$ (Fig. ~\ref{fig:diagrams}b). Logarithmic divergences always appear in the hole-hole channel when $\bm{p}+\bm{p}'=0$ as is usual for Fermi liquids, however in the nested situation, additional logarithmic divergences appear when $\bm{p}$ and $\bm{p}'$ lie on opposite nested sides and $\bm{p}+\bm{p}'\neq 0$. In the hole-electron channel, logarithmic divergences appear only when nesting is present, with $\bm{p}$, $\bm{p}'$ required to lie on opposite nested sides of the Fermi surface.

Virtual processes shown in Fig. ~\ref{fig:diagrams}a provide a negative correction to the forward scattering vertex, $\delta \Gamma^\Lambda_{\sigma_1\sigma_2\sigma_1\sigma_2}(\bm{k}_1,\bm{k}_2,\bm{k}_1,\bm{k}_2) = -\sum_{\bm{p};\sigma,\sigma'}{
|\Gamma^\Lambda_{\sigma_1\sigma_2\sigma\sigma'}(\bm{k}_1,\bm{k}_2,\bm{p},\bm{p}')|^2/|E_{hh}|}$ where momentum conservation implies $\bm{p}'=\bm{k}_1+\bm{k}_2-\bm{p}$, and $E_{hh}$ is the energy of the intermediate two-hole pair. Thus, for the repulsive hole-hole interactions which we consider, this diagram provides a suppression of forward scattering. For the process shown in Fig. ~\ref{fig:diagrams}b, the correction is always positive, $\delta \Gamma^\Lambda_{\sigma_1\sigma_2\sigma_1\sigma_2}(\bm{k}_1,\bm{k}_2,\bm{k}_1,\bm{k}_2) = +\sum_{\bm{p};\sigma,\sigma'}{
|\Gamma^\Lambda_{\sigma_1\sigma \sigma'\sigma_2}(\bm{k}_1,\bm{p},\bm{p}',\bm{k}_2)|^2/|E_{he}|}$ where  momentum conservation implies $ \bm{p}'=\bm{p}+ \bm{k}_2-\bm{k}_1$ and $E_{he}$ is the energy of the hole-electron pair; thus these virtual processes enhance forward scattering.

\begin{figure}
\includegraphics[width = 0.4\textwidth]{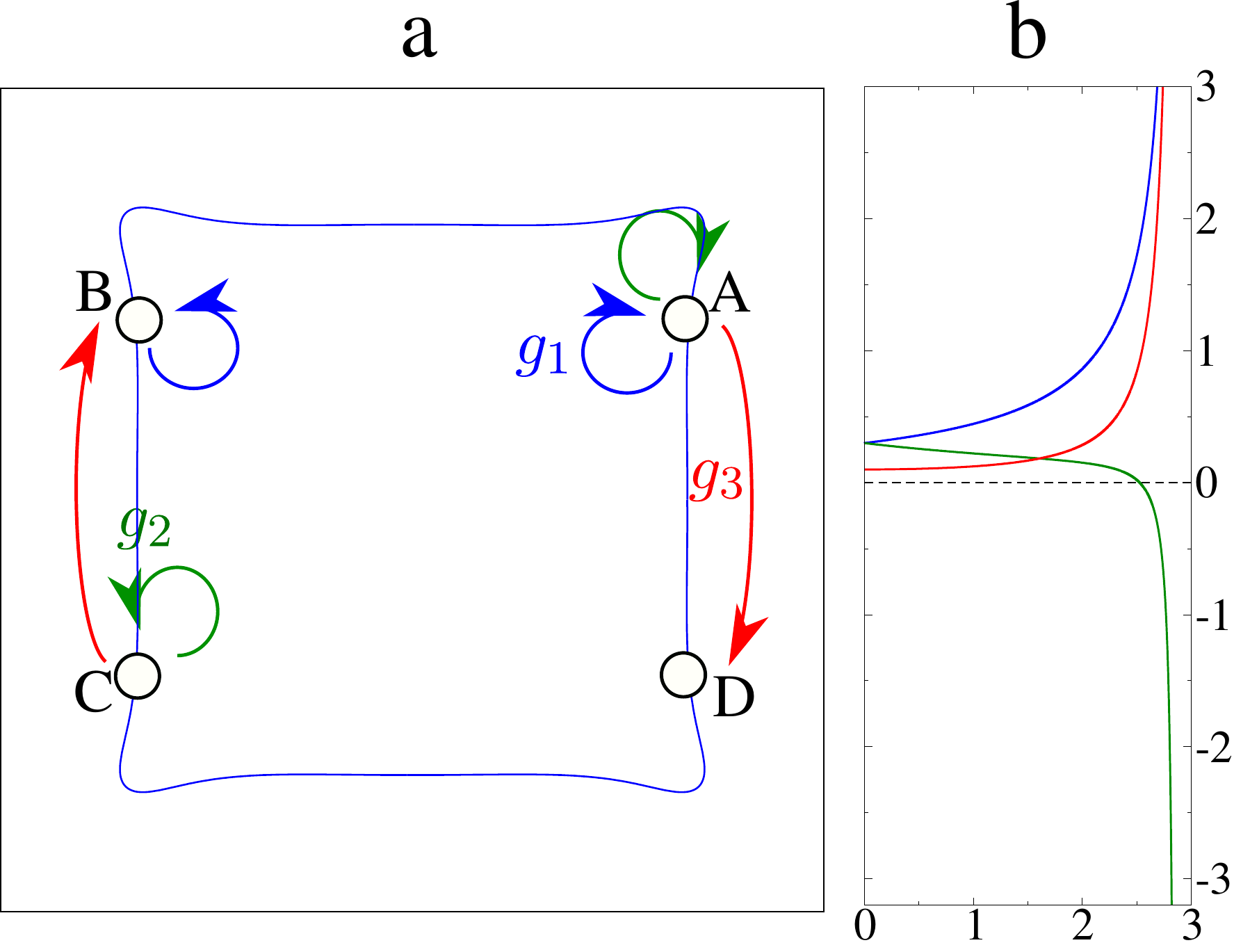} 
\caption{(a) Coupling constants used in the patch RG model $g_1 = \nu\Gamma(ABAB)$, $g_2 = \nu\Gamma(ACAC)$, $g_3 = \nu\Gamma(ACDB)$. (b) RG flow of the couplings $g_1,g_2,g_3$ plotted in the blue, green and red lines respectively, for initial values $g_1=g_2 = 0.3, g_3=0.1$.}
\label{fig:scattering}
\end{figure}

For general momenta $\bm{k}_1,\bm{k}_2$, both diagrams in Fig. ~\ref{fig:diagrams} contribute, however a simple phase space argument shows that, at special points on the Fermi surface, the contribution from Fig. ~\ref{fig:diagrams}a vanishes, resulting in a positive definite RG flow and therefore a divergence in the running coupling at a finite energy scale, indicating an instability. These scattering processes correspond to $\bm{k}_1,\bm{k}_2$ lying laterally opposite at the edges of parallel nested portions of the Fermi surface, labelled A,B in Fig. ~\ref{fig:scattering}. Since $\bm{k}_1+\bm{k}_2\neq 0$, the diagram Fig. ~\ref{fig:diagrams}a diverges only if the internal momenta $\bm{p},\bm{p}'$ lie on nested regions of the Fermi surface. Choosing the components of both $\bm{k}_1$ and $\bm{k}_2$ parallel to the Fermi surface to be equal to the maximum possible values for momenta lying in the nested regions, the momentum conservation condition $\bm{k}_1+\bm{k}_2=\bm{p}+\bm{p}'$ can only be satisfied for a vanishingly small momentum space volume, and there is no contribution to the RG flow. 

At the same time, we note that the contribution to the RG flow from Fig. ~\ref{fig:diagrams}b arises from processes in which $\bm{p}$ can take any value along the nested portion of the Fermi surface, since $\bm{k}_2-\bm{k}_1 =\bm{Q}$ is a nesting vector and $\bm{p}'=\bm{p}+\bm{Q}$ lies laterally opposite to $\bm{p}$ for any value of $\bm{p}$ on a straight section of the Fermi surface. Since Fig. ~\ref{fig:diagrams}b always provides a positive correction and never vanishes, the coupling grows to $+\infty$ under the RG flow. This allows us to establish that the Fermi liquid phase is unstable, with one possible instability involving the condensation of hole-electron pairs with total momentum $\bm{Q}$, which gives rise to a density wave.

While we have established an instability via generic phase space volume arguments, determining whether this instability is dominant requires analysis of a more detailed and specific model. One alternative to the density wave instability is superconductivity, which is driven by interactions for which the corrections due to virtual scattering in the hole-hole channel (Fig. ~\ref{fig:diagrams}a) rather than the hole-electron channel (Fig. ~\ref{fig:diagrams}b) dominate. Generic phase space arguments allow us to identify one such interaction, which involves forward scattering in the Cooper channel, $\bm{k}_1+\bm{k}_2 = 0$, with $\bm{k}_1,\bm{k}_2$ lying at the edges of the nested portions of the Fermi surface (the points labelled A,C in Fig. ~\ref{fig:scattering}). The logarithmically divergent processes corresponding to Fig. ~\ref{fig:diagrams}a involve hole-hole pairs with $\bm{p}'=-\bm{p}$; when $\bm{p}$ lies in one nested region, $\bm{p}'$ automatically lies on the opposite nested region, thus the phase space volume is maximized. However, the hole-electron processes in Fig. ~\ref{fig:diagrams}b provide a vanishing contribution to the RG flow, since $\bm{k}_2-\bm{k}_1$ is a vector of maximum length connecting the nested regions of opposite sides of the Fermi surface, and the momentum conservation condition $\bm{p}'-\bm{p} = \bm{k}_2-\bm{k}_1$ with $\bm{p},\bm{p}'$ in the nested regions can only be fulfilled within a vanishingly small momentum space volume. Thus the correction is purely due to Fig. ~\ref{fig:diagrams}a which is negative. If the correction itself vanishes at low energies, then the interaction flows to zero; however if the correction remains finite, the interaction will grow to $-\infty$ and lead to a superconducting instability. The former situation occurs ordinarily in Fermi liquids, however, when nesting is present, the second situation is generically true, due to the fact that the individual effective vertices appearing in the second-order processes are themselves dressed by both virtual hole-hole and hole-electron processes.

We may introduce a minimal model possessing both density wave and superconducting instabilities, involving only one pair of nested sides, shown in Fig. ~\ref{fig:scattering}a. We define interactions on four patches of the Fermi surface close to the points A,B,C,D; we find that in these regions the velocity normal of the Fermi surface is minimum, which enhances the interactions. These patches play a similar role to the ``hot-spots" in previously studied patch RG models of the cuprates \cite{Schulz1987,Dzyaloshinskii1987,Furukawa1998}. Rather than coupling functions, we introduce a discrete set of coupling constants $g_1 = \nu\Gamma_{\sigma\sigma\sigma\sigma}(ABAB), g_2 = \nu\Gamma_{\sigma\sigma\sigma\sigma}(ACAC), g_3 = \nu\Gamma_{\sigma\sigma\sigma\sigma}(ACDB)$, neglecting backscattering and spin-flip scattering, with $\nu$ being the single-band density of states averaged over a single patch. The scaling relations are
\begin{gather}
\frac{dg_1}{dl} = g_1^2 + g_3^2  \ \ , \nonumber \\
\frac{dg_2}{dl} = -g_2^2 -g_3^2  \ \ , \nonumber \\
\frac{dg_3}{dl} =2 g_1g_3 - 2g_2g_3 \ \ .
\end{gather}
For an initial value $g_3(l=0) = 0$, we find that $g_1$ diverges at a scale $l = \log(\Lambda_0/\Lambda) = 1/g_1$, while $g_2 \rightarrow 0$. However, when $g_3(l=0)\neq 0$, we find that $g_3$ is monotonically enhanced by the RG flow, so that $dg_2/dl < -g_3^2$ is strictly negative, leading to $g_2 \rightarrow -\infty$. The competition between the density wave and superconducting phases may be resolved by determining which coupling diverges the fastest, since the RG flow must be stopped when the couplings become large. In Fig. ~\ref{fig:scattering}b we plot the flow of the couplings for initial values $g_1=g_2=0.3, g_3=0.1$. We find that, since $g_2$ must first decrease from an initially positive value, it reaches strong coupling at a later RG time than $g_1$ and thus the dominant instability is associated with $g_1$, corresponding to a $\bm{Q}$-density wave.

While this model is illustrative, it is not predictive since it employs a very small subset of the possible couplings that contribute to the RG flow. The most significant modification to the scaling relations arises from scattering in the Cooper channel ($\bm{k}_1+\bm{k}_2 =\bm{k}_3+\bm{k}_4=0$) involving large momentum transfer, which enhances the flow of $g_2$. While the couplings are initially small, they may grow sufficiently rapidly under the RG flow to allow $g_2$ to diverge faster than $g_1$, in which case the superconducting instability becomes dominant.

\begin{figure*}[t]
\includegraphics[width = 0.6\textwidth]{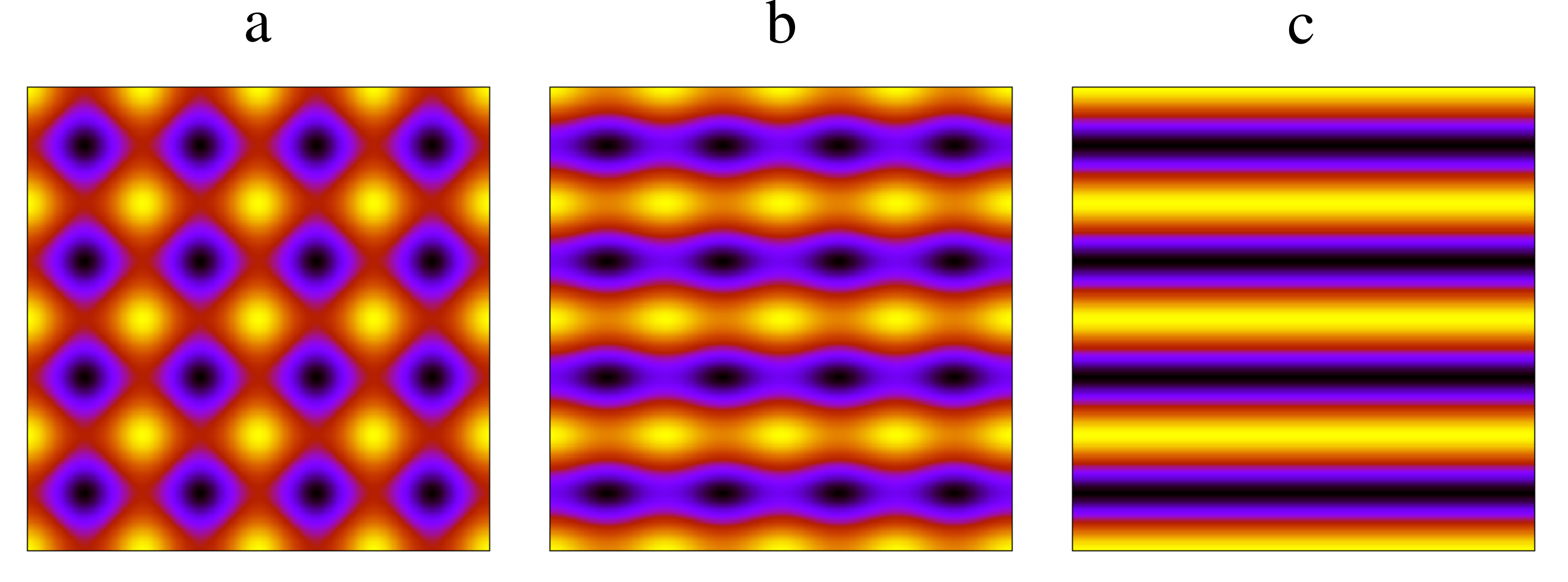} 
\caption{The spatial modulation of the charge density in the CDW phase for the cases where (a) four sides of the Fermi surface exhibit nesting equally, (b)  the left and right sides are nested over a greater region than the top and bottom sides, and (c) only the left and right sides are nested. Darker/brighter regions correspond to higher/lower density.}
\label{fig:CDW}
\end{figure*}

\section{Properties of the ordered phases}

When the Fermi surfaces are non-degenerate, as is the case when either a magnetic field is present or bulk inversion symmetry is broken, the RG analysis predicts a CDW ground state, corresponding to the condensation of hole-electron pairs on opposite nested sides of the Fermi surface.  In the absence of a magnetic field, there are two perpendicular nesting vectors $\bm{Q}_1,\bm{Q}_2$ along the $\hat{\bm{x}},\hat{\bm{y}}$ directions, and the CDW order forms a 2D checkerboard lattice. However, when a magnetic field is present, nesting can be significantly suppressed or destroyed for one pair of sides of the Fermi surface, leading to a striped CDW order. We illustrate these cases in Fig. ~\ref{fig:CDW}.

In the case of degenerate Fermi surfaces, both CDW and SDW ordering is possible. In order to determine the ground state, we must minimize the Landau-Ginzburg free energy simultaneously with respect to the CDW and SDW order parameters, which we may collect into a four-dimensional complex vector $\bm{\Phi} = (\varphi,M_x,M_y,M_z)$ with $\varphi$ being the CDW order parameter and $(M_x,M_y,M_z)$ the SDW order parameter. The free energy may be calculated by performing a Hubbard-Stratonovich transformation to decompose the hole-hole interaction into a coupling between holes and the bosonic order parameters. We divide the action into contributions involving the left/right and top/bottom sides of the Fermi surface $\mathcal{S}_{LR} + \mathcal{S}_{TB}$, and accounting for the $g_1$ interaction the Euclidean action for one pair of sides is
\begin{gather}
\mathcal{S}_{LR} = \int{}d\tau \{\frac{1}{g_1} \bm{\Phi}^* \cdot\bm{\Phi}+ \nonumber \\ \sum_{\bm{k} \in R, L}{c^\dag_{\bm{k}}\left[\partial_\tau + \varepsilon_{\bm{k}}\right] c_{\bm{k}}}
+ \sum_{\bm{k}\in L}{\left[
c^\dag_{\bm{k}+\bm{Q}} \bm{\Phi}\cdot \bm{\sigma} c_{\bm{k}} + \text{H.c.}\right]}\}
\end{gather}
where $c^\dag_{\bm{k}} = (c^\dag_{\bm{k}+},c^\dag_{\bm{k}-})$ is a two-component spinor and $\bm{\sigma} = (\sigma_0,\sigma_x,\sigma_y,\sigma_z)$ with $\sigma_0$ being the identity matrix.
Integration over the fermionic degrees of freedom yields the Landau-Ginzburg free energy
\begin{gather}
\mathcal{F} = \left(g_1^{-1} - a(T)\right)|(|\varphi|^2 + |\bm{M}|^2)
+ \nonumber \\
b(T)\{
|\varphi|^4 + |\bm{M}|^4 + |\bm{M}\times\bm{M}^*|^2 + 4|\varphi|^2|\bm{M}|^2 + \nonumber \\ (\varphi^*)^2\bm{M}\cdot\bm{M}
+\varphi^2 \bm{M}^*\cdot\bm{M}^* \}
\end{gather}
where we have ignored fluctuations of the order parameters, and
\begin{gather}
a(T) = \frac{1}{2}\sum_{\omega_n,\bm{p}\in L}{\frac{1}{\omega_n^2 + \varepsilon^2_{\bm{p}}}} \ ,  \ b(T) = \frac{1}{16}\sum_{\omega_n,\bm{p}\in L}{\frac{1}{(\omega_n^2 + \varepsilon_{\bm{p}}^2)^2}} \ ,
\end{gather}
with $\omega_n$ being the fermionic Matsubara frequencies.

Minimizing the free energy yields degenerate CDW and SDW orders, with $\varphi$ purely real and $\bm{M}$ purely imaginary, thus $(|\varphi|,\bm{M})$ is a real four-dimensional vector. The free energy possesses an SO(4) symmetry and the system will spontaneously choose any orientation in the four-dimensional order parameter manifold. However, this SO(4) symmetry is an artifact of our restricted model in which only the $g_1$ coupling, which we assume dominates at low energies, is present. The symmetry is broken as soon as subleading interactions are considered. The most significant effect is due to the spin dependence of the interactions, which are negligible for small angle scattering but substantial for interactions involving large momentum transfer.

An accurate treatment of the density wave orders accounting for the large range of possible couplings is beyond the scope of this work, however, we may form some generic conclusions by considering the spin structure of the interactions, which is restricted by the presence of both inversion and time reversal symmetry (since we are considering the case when the Fermi surfaces are spin degenerate). In order to do this we define a basis of single-hole states $|\bm{k},\sigma= \pm\rangle$ which are eigenstates of the combined symmetry operation $IC_{2z}$ consisting of inversion and a twofold combined spin and orbital rotation about the $z$--axis with eigenvalue  $e^{-\frac{3i\pi}{2}}\sigma$. States with opposite $\sigma$ are related by the combination of time reversal and spatial inversion, $|\bm{k},-\rangle = I\mathcal{T}|\bm{k},+\rangle$. The symmetries of the Hamiltonian then imply that $\sigma_1\sigma_2=\sigma_3\sigma_4$, which restricts the possible couplings to $\Gamma_{++++}=\Gamma_{----}^*$, $\Gamma_{+-+-} = \Gamma_{-+-+}^*$,  $\Gamma_{+--+} = \Gamma^*_{-++-}$, $\Gamma_{++--} = \Gamma^*_{--++}$. Performing a Hubbard-Stratonovich decomposition of these interactions leads to additional terms in the free energy containing products $\delta F = \Gamma_{0z}\varphi^* M_z+\Gamma_{xy} M_x^* M_y$ and their complex conjugates. Noting that the quartic terms in the only contain even powers of $\varphi$, the derivative of the quartic terms with respect to $\varphi$ vanish at $\varphi = 0$, and thus
\begin{gather}
\frac{\partial \mathcal{F}}{\partial \varphi^*}|_{\varphi = 0} = \Gamma_{0z} M_z  \ \ ,
\end{gather}
which implies that, at the minimum of the free energy, $\varphi$  cannot vanish unless $M_z$ also vanishes. Therefore the ground state can either consist of an in-plane SDW ($\varphi = M_z=0$), or coexisting CDW and out-of-plane SDW orders ($\varphi,M_z\neq 0$).

The presence of CDW/SDW order leads to a Fermi surface reconstruction which bears similarities to that observed in the cuprates \cite{Norman1998}. While the nested portions of the Fermi surface are gapped by density wave ordering, the corners do not couple to the order parameter and maintain a nonvanishing density of states at the Fermi level. The reconstructed Fermi surface consists of Fermi arcs formed from the unnested corner regions of the original Fermi surface.

\section{Conclusion}

We have demonstrated the existence of a generic nesting phenomenon in semiconductor quantum wells formed in materials with a cubic lattice structure, resulting in density wave instabilities, although our analysis does not exclude the possibility of a competing superconducting ground state. In the ordered phases we explore, the single-electron spectrum is partially gapped, but the corner regions of the Fermi surface which are not nested remain ungapped, allowing the ground state to remain metallic. The correlated ground state bears striking similarities to the pseudogap phase observed in the cuprates, which have been observed to exhibit Fermi arcs as well as density wave order \cite{Norman1998,Vershinin2004,Comin2014}.

The realization of these features in semiconductor systems, which possess extremely long mean free paths (of the order of microns \cite{Dobbie2012}), presents an important opportunity for further study to determine whether these systems share other physical properties with the cuprates. The advantages of using highly versatile nanofabricated semiconductor systems over high-$T_c$ superconductors include the ability to precisely tune the density \emph{in situ} via gates, which combined with detailed knowledge of the band structure provides a setting in which the source of the interaction effects may be definitively known, unlike in high-$T_c$ superconductors, in which the role of nesting is still not conclusively established. Since the dimensionless interactions depend on the velocity normal to the Fermi surface at the ``hot-spots" (the points A,B,C,D indicated in Fig. ~\ref{fig:scattering}), which is controlled by the width of the well $v_\perp \propto 1/d$, a crucial opportunity exists to control the interaction strength.

The nesting vector is determined by the width of the quantum well $|\bm{Q}|\sim 2\pi/d$ rather than the lattice spacing, and the ordered phases exhibit a continuous symmetry breaking and gapless sliding modes which might enhance the role of fluctuations. Novel interactions between holes may emerge, enriching the phenomenology of the correlated ground state and allowing further comparisons with the pseudogap phase in the high-$T_c$ superconductors. Our systems are also expected to exhibit a higher degree of nesting than those possible in the cuprates as well as other interacting systems including graphene \cite{Nandkishore2012}, kagome materials \cite{Ortiz2020,Zhu2021,Chen2021,Ortiz2021,Ni2021,Chenb2021,Liang2021} and Moir\'{e} superlattices \cite{Wang2020,Jin2021,Xu2020,Huang2021,Miao2021,Ghiotto2021,Shi2021}, since in our case exact nesting emerges upon tuning of the density to a transition from two distinct Fermi geometries which always exist for doping in different regimes. The existence of a nesting density is thus guaranteed in real systems across a wide range of materials. This is in contrast to nested Fermi surfaces appearing in lattice models, in which nesting can be destroyed e.g. by next-nearest-neighbor hopping in the case of the square lattice and third neighbor hopping for honeycomb lattices.

In our analysis, realization of nested Fermi surfaces requires a number of specific conditions: firstly, the cubic symmetry of the lattice must be maintained in the heterostructure. This implies that the crystal growth axis must be aligned along a high-symmetry direction. Furthermore, while for Ge and Si quantum wells, the asymmetry of the transverse confining potential does not prohibit the existence of nested Fermi surfaces, for zincblende materials such as GaAs, nesting is completely destroyed when the asymmetry is significant (see Fig. ~\ref{sfig:Rashba} in the Supplementary Material).

While we have used the perturbative RG method to demonstrate the existence of ordered phases for weak coupling, our analysis suffers from the common issue among 2D systems that the couplings may realistically be too large for perturbative methods to be justified. Our analysis is applicable only for the case when the superconducting (SC)/CDW/SDW gaps are exponentially smaller than an ultraviolet scale $\Lambda_0$ which is nominally similar to the Fermi energy, which at the nesting density is $\approx 5$ meV in Ge and $\approx 2.5$ meV in GaAs; the gap is $\Delta \sim \Lambda_0 e^{-l}$ where $l \gg 1$ is the RG scale at which the flows enter the strong coupling regime, with the bare couplings being $\approx l^{-1}$. We have not attempted to determine a value for $\Delta$, since the precise values of the bare interactions that enter into our weak-coupling RG flow cannot be accurately predicted. We have included terms in the scaling relations which contain the logarithmically divergent contributions to the interaction vertex arising from nesting, neglecting RPA screening effects as well as vertex corrections, which should enter at the ultraviolet scale and present a significant modification of the initial values entering the RG flow. Variation of the well width provides the possibility of varying the interactions between the weak and strong coupling regimes, which may reveal a more complex phase diagram than the one we have established via perturbative methods.

Disorder also poses an important issue for the realization of the ordered ground states predicted by our RG analysis. In order for the correlated phases to be observable, the SC/CDW/SDW gaps must be larger than the variation of the chemical potential due to disorder, i.e. $\tau\Delta  = \tau\varepsilon_F  e^{-l}>1$. The typical mobilities $\mu\approx 10^6$ cm$^{2}$V$^{-1}$s$^{-1}$ observed at densities $p \approx 10^{11}$cm$^{-2}$ imply a disorder lifetime $\varepsilon_F \tau \approx 200$ in typical experiments \cite{Dobbie2012}. Assuming that $\varepsilon_F \tau$ takes similar values for the higher dopings required to observe nesting (which require an increase in the Fermi energy by approximately a factor of two), we would expect the correlated phases to be observable unless the dimensionless initial couplings are very small, $\tau\Delta < 1\rightarrow l^{-1} < 0.2$.

We also need to the address the question of whether the nesting densities are attainable in real systems. While we have related nesting to a transition in the Fermi surface geometry which occurs at a single density, we have also seen that the Fermi surfaces exhibit perfect nesting over an extended range of dopings (see Fig. ~\ref{fig:FS_Ge}b), thus extreme fine tuning is not required. The nesting density predicted for GaAs lies within the range currently used for experiments \cite{Marcellina2018}. This leaves open the possibility that the correlated phases have already been created, although so far there have been no specific studies to identify them.

\section{Acknowledgements}

H. D. Scammell acknowledges funding from ARC Centre of Excellence FLEET. TL acknowledges support from the Deutsche Forschungsgemeinschaft.

\widetext
\newpage

\begin{center}
\textbf{\large Supplementary Figures}
\end{center}

\setcounter{equation}{0}
\setcounter{figure}{0}
\setcounter{table}{0}
\setcounter{page}{1}
\makeatletter
\renewcommand{\theequation}{S\arabic{equation}}
\renewcommand{\thefigure}{S\arabic{figure}}
\renewcommand{\bibnumfmt}[1]{[S#1]}
\renewcommand{\citenumfont}[1]{S#1}

\begin{figure}[h]
\includegraphics[width = 0.75\textwidth]{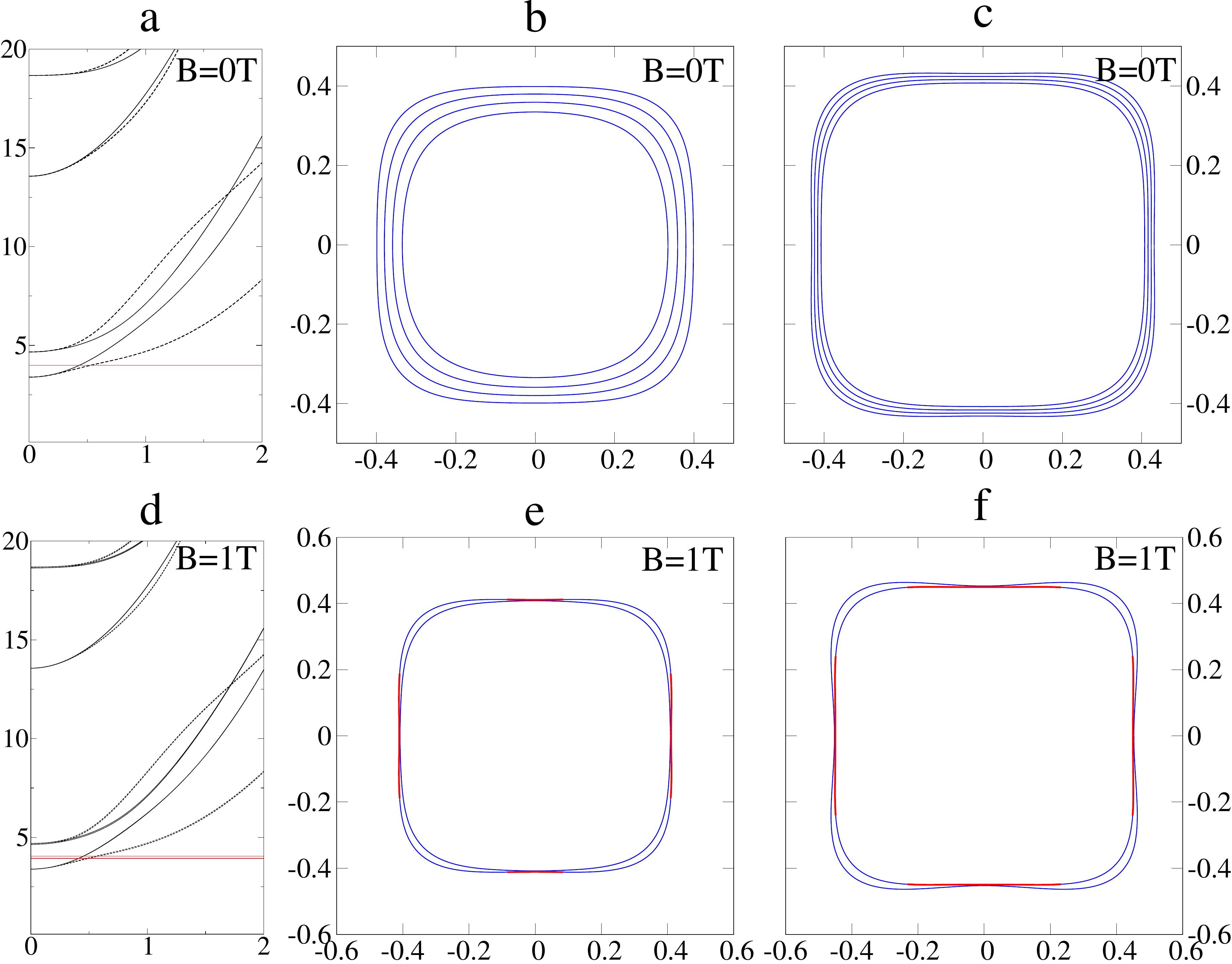} 
\caption{\textbf{Nesting in Si wells}. Evolution of the Fermi surface with doping in a Si quantum well with infinite square well confinement, in zero magnetic field (a,b,c) and a 1T magnetic field along (010) (d,e,f). (a,d) The 2D dispersion (in meV) for a $d=20$ quantum well. The red lines indicate the Fermi energy at the nesting densities. (b,c,e,f): Fermi surfaces at various hole densities (in units of $10^{10}$ cm$^{-2}$) (b): $p = 4.83, 5.67, 6.46, 7.27$, (c): $p = 7.67, 8.08, 8.47, 8.86$, (e) $p=5.49$, (f) $p=6.28$. Portions of the Fermi surface on which the normal momentum coordinate varies by less than 0.25\% are highlighted in red. Units of wavevector are $\pi/d$.}
\label{sfig:Si}
\end{figure}

\begin{figure}[h]
\includegraphics[width =0.75\textwidth]{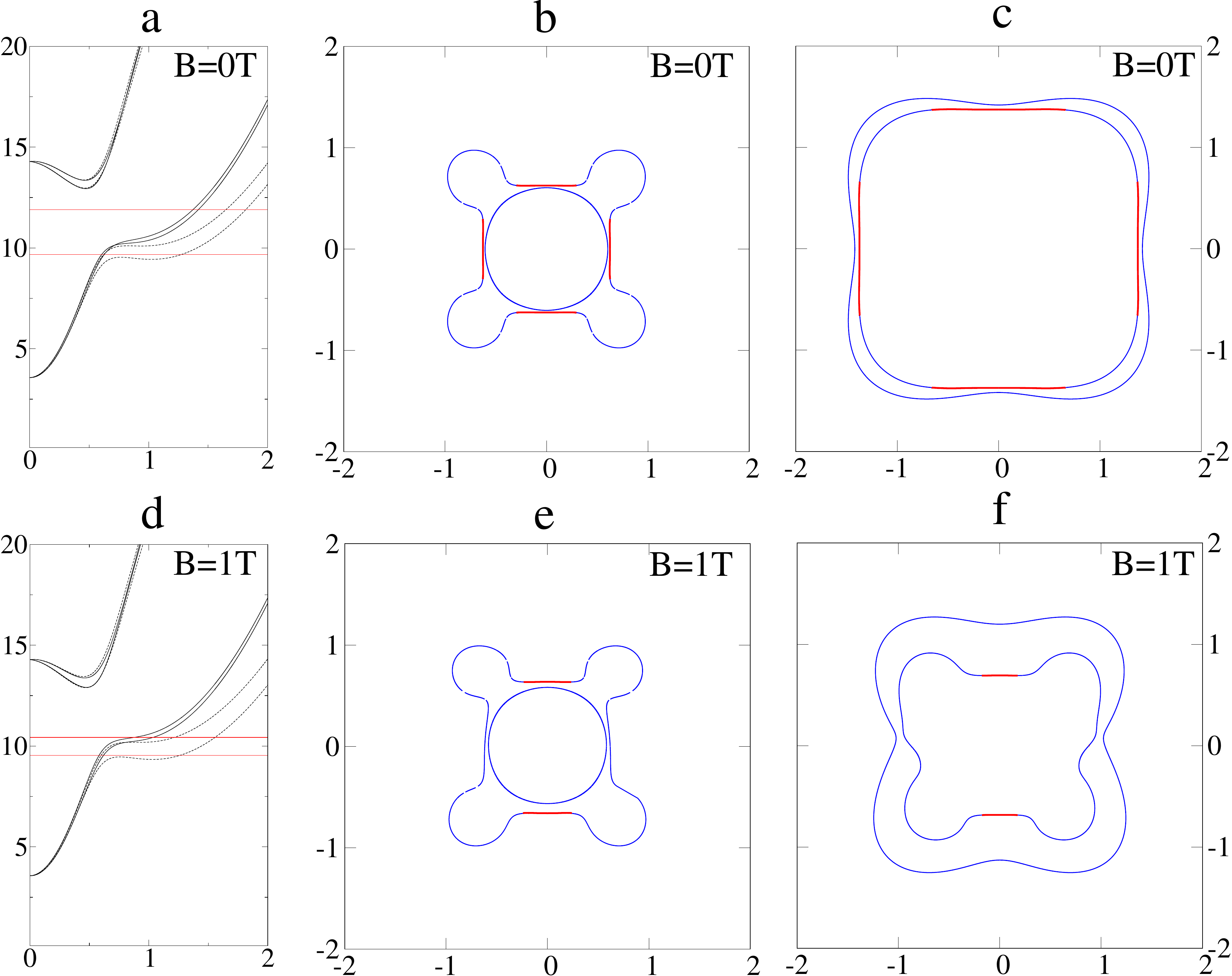}
\caption{\textbf{Nesting in InAs wells}. Nested Fermi surfaces in an InAs quantum well with infinite square well confinement, in zero magnetic field (a,b,c) and a 1T magnetic field along (010) (d,e,f). (a,d) The 2D dispersion (in meV) for a $d=20$ quantum well. The red lines indicate the Fermi energy at the nesting densities. (b,c,e,f): The Fermi surface for various hole densities (b) $p = 2.64\times 10^{11}$ cm$^{-2}$ (c) $p = 9.56\times 10^{11}$, (e) $p = 2.67\times 10^{11}$ cm$^{-2}$ and (f) $p = 5.37\times 10^{11}$ cm$^{-2}$. Portions of the Fermi surface on which the normal momentum coordinate varies by less than 0.25\% are highlighted in red. Units of wavevector are $\pi/d$.}
\label{sfig:InAs}
\end{figure}

\begin{figure}[h]
\includegraphics[width =0.5\textwidth]{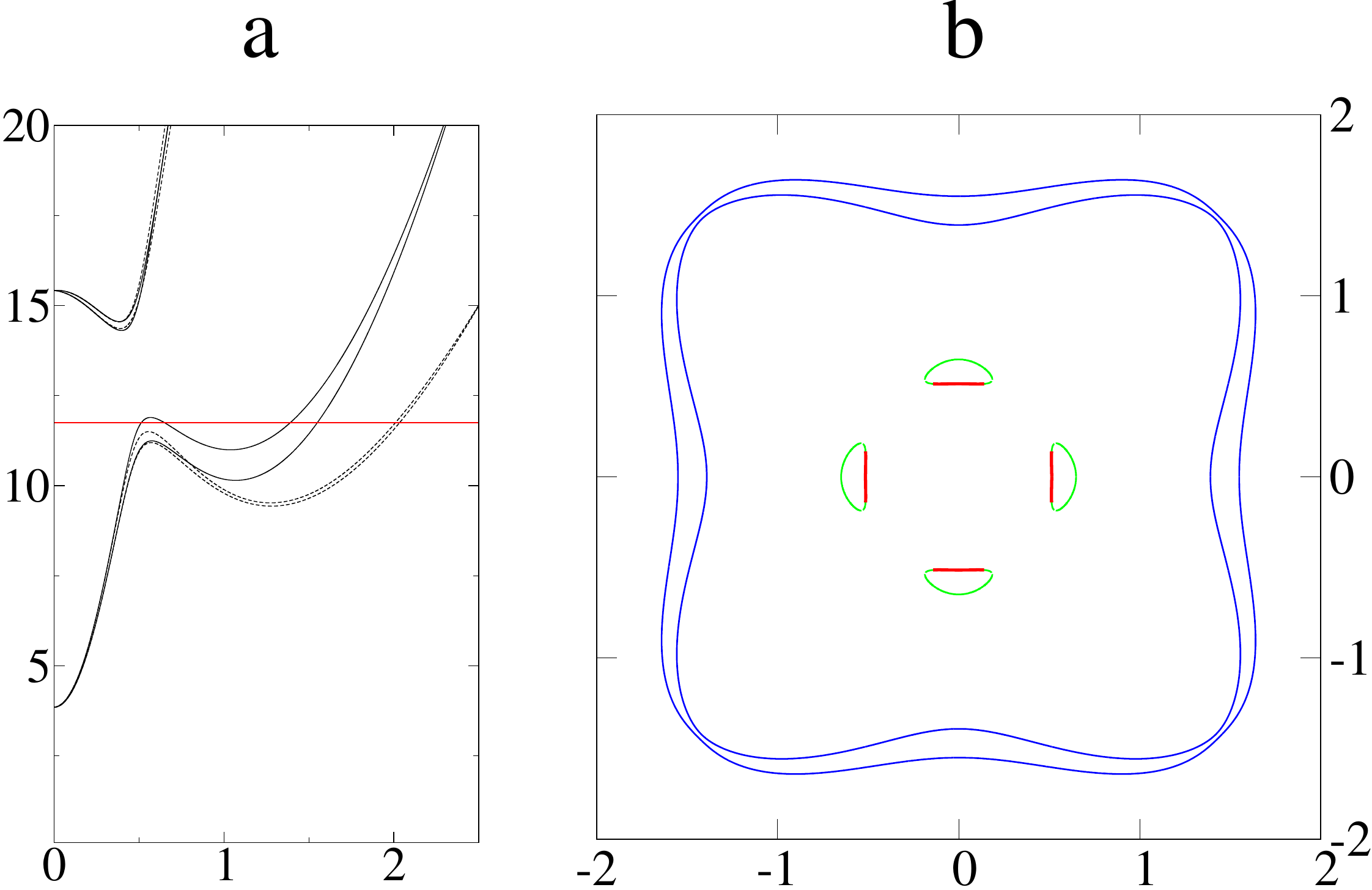}
\caption{\textbf{Nesting in InSb wells}. Nested Fermi surfaces in an InSb quantum well with infinite square well confinement in zero magnetic field. (a) The 2D dispersion (in meV) for a $d=20$ quantum well. The red lines indicate the Fermi energy at the nesting densities. (b) The Fermi surface at the hole density $p= 1.42\times 10^{12}$ cm$^{-2}$. In addition to two large hole Fermi surfaces, four small electron pockets exist in the upper band. Nesting occurs between the electron pockets rather than across the exterior sides of the hole Fermi surfaces.  The unnested portions of the boundaries of the electron pockets are shown in green. No nesting densities were found for a $B=$1T magnetic field along (010), unlike in the other materials discussed in the Supplementary Material. Portions of the Fermi surface on which the normal momentum coordinate varies by less than 0.25\% are highlighted in red. Units of wavevector are $\pi/d$.}
\label{sfig:InSb}
\end{figure}

\begin{figure}[h]
\includegraphics[width = 0.8\textwidth]{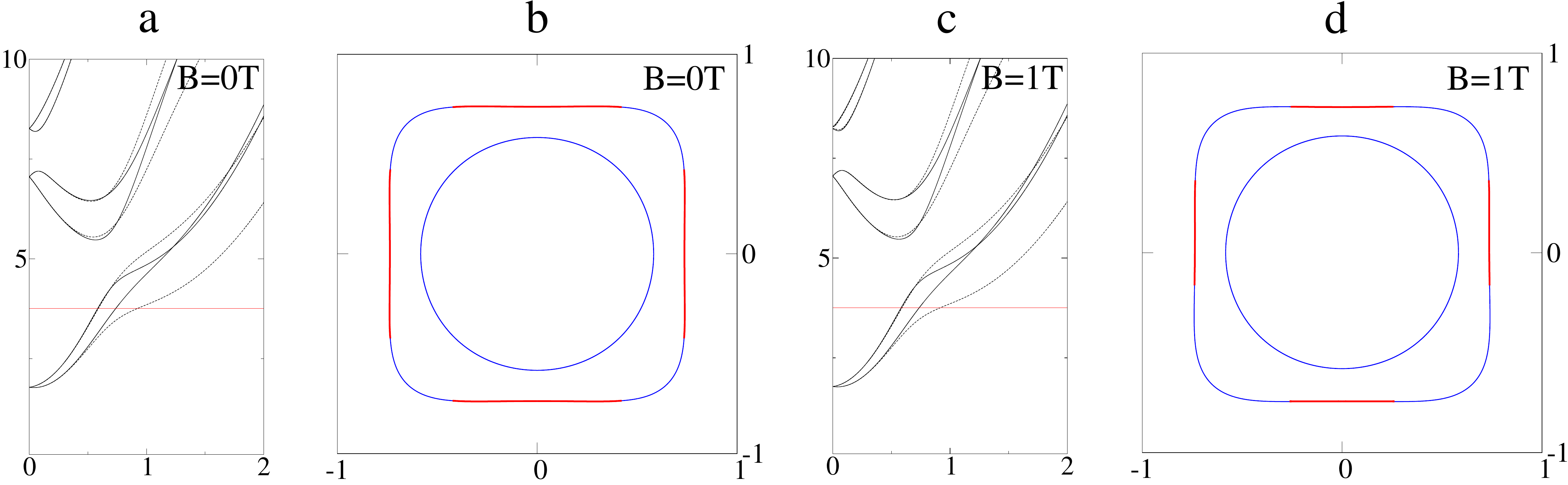}
\caption{\textbf{Nesting in CdTe wells}. Nested Fermi surfaces in a CdTe quantum well with infinite square well confinement, in zero magnetic field (a,b) and a 1T magnetic field along (010) (c,d). (a,c) The 2D dispersion (in meV) for a $d=20$ quantum well. The red lines indicate the Fermi energy at the nesting densities. (b,d) The Fermi surface for the hole density  $p = 1.98\times 10^{11}$ cm$^{-2}$. Portions of the Fermi surface on which the normal momentum coordinate varies by less than 0.25\% are highlighted in red. Units of wavevector are $\pi/d$. Unlike the other materials studied, no densities were found at which the smaller Fermi surface was nested. This may be explained by the fact that, for $\bm{k}$ along the (100) and (010) directions, the curvature $v_\perp^{-1}\partial^2\varepsilon/\partial k_\parallel^2$ (where $v_\perp$ and $k_\parallel$ are the velocity perpendicular and momentum parallel to the equal-energy contours) never vanishes, as can be seen in Fig. ~\ref{sfig:concavity}.}
\label{sfig:CdTe}
\end{figure}

\begin{figure}[h]
\includegraphics[width =0.75\textwidth]{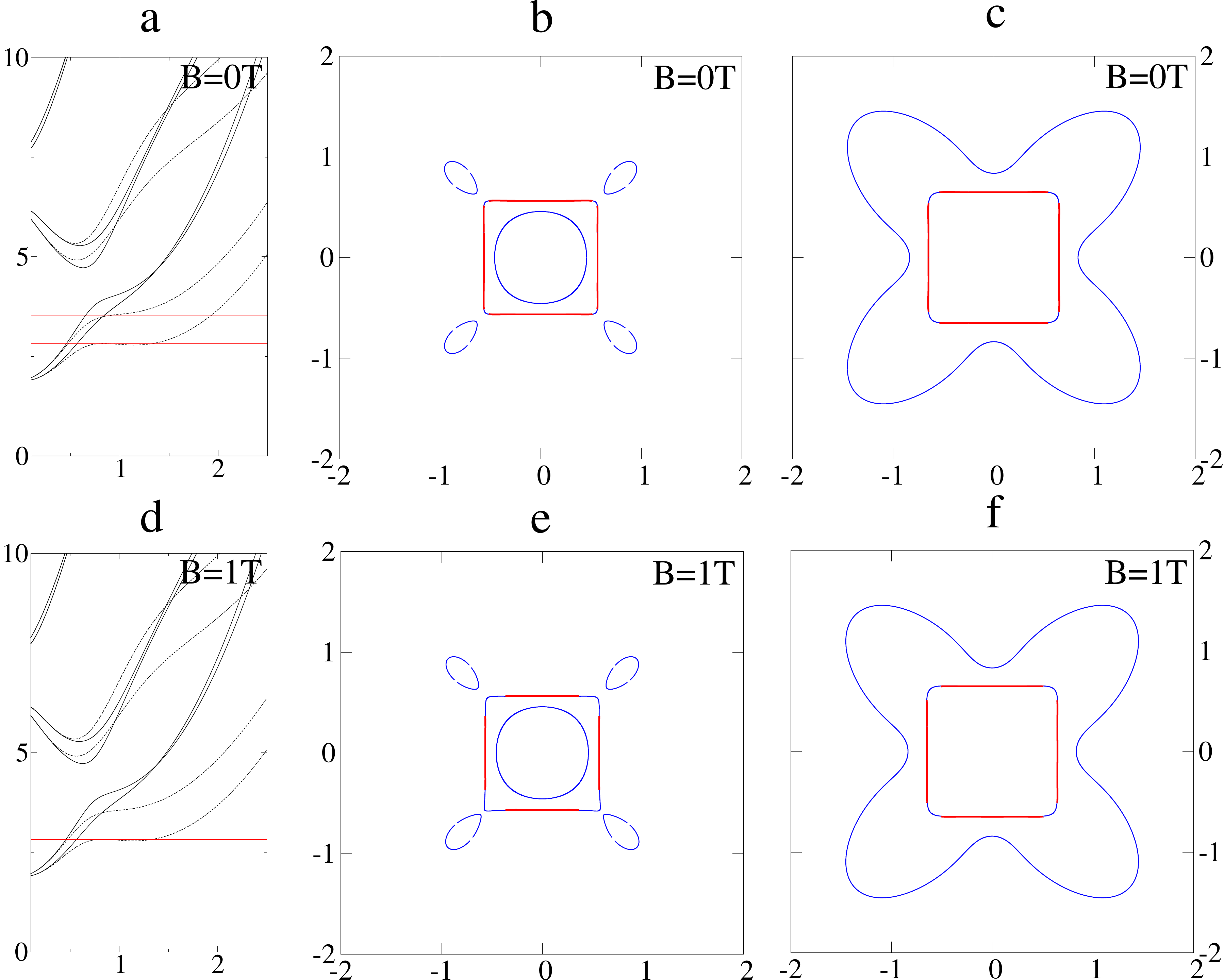}
\caption{\textbf{Nesting in ZnSe wells}.
Nested Fermi surfaces in a ZnSe quantum well with infinite square well confinement, in zero magnetic field (a,b,c) and a 1T magnetic field along (010) (d,e,f). (a,d) The 2D dispersion (in meV) for a $d=20$ quantum well. The red lines indicate the Fermi energy at the nesting densities. (b,c,e,f): The Fermi surface for various hole densities (b) $1.53\times 10^{11}$ cm$^{-2}$ (c) $4.88\times 10^{11}$ cm$^{-2}$, (e) $p=1.52\times 10^{11}$ cm$^{-2}$ and (f) $4.86\times 10^{11}$ cm$^{-2}$. Portions of the Fermi surface on which the normal momentum coordinate varies by less than 0.25\% are highlighted in red. Units of wavevector are $\pi/d$.}
\label{sfig:ZnSe}
\end{figure}

\begin{figure}[h]
\includegraphics[width = 0.7\textwidth]{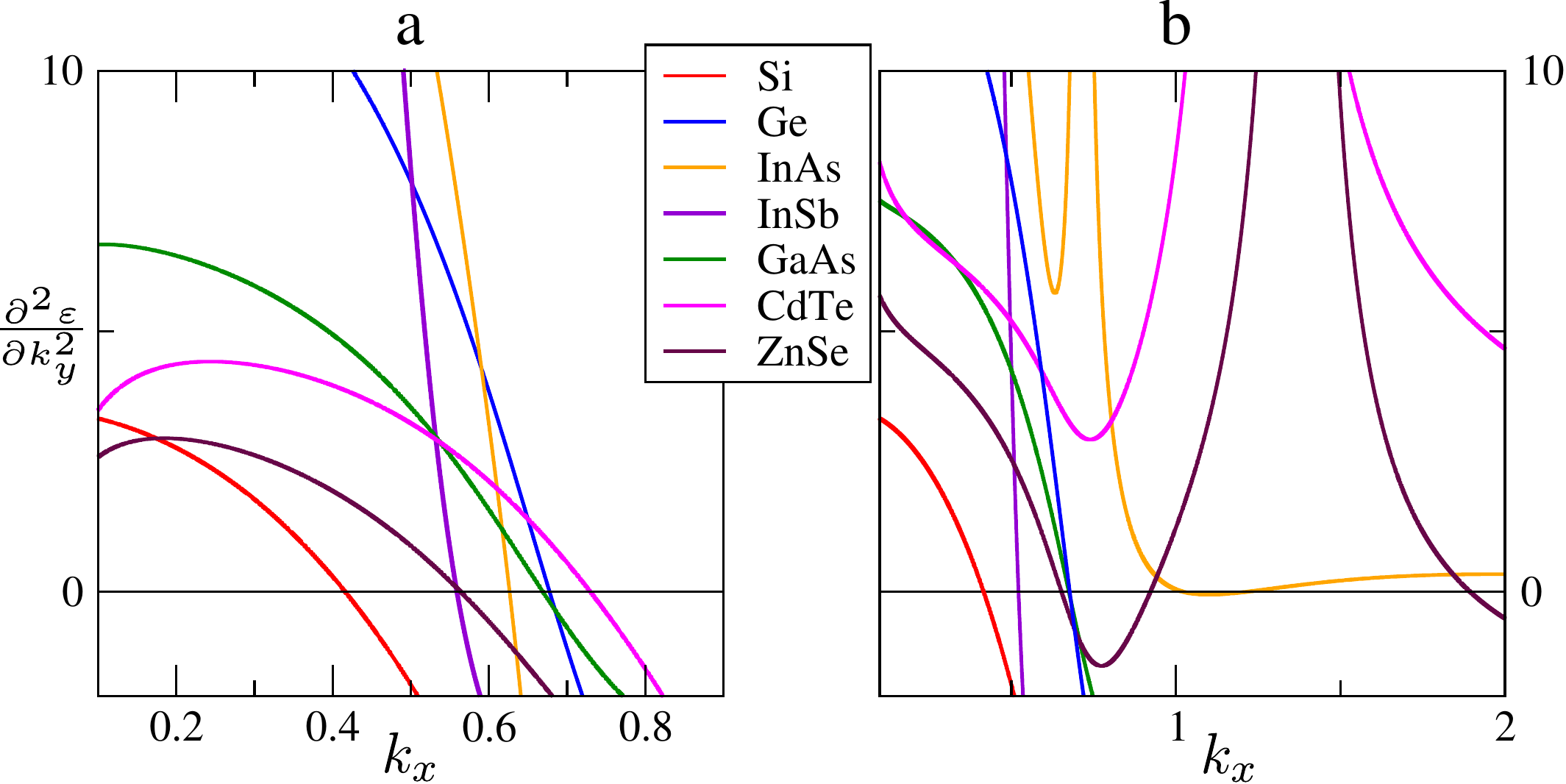}
\caption{\textbf{Origin of nesting in cubic semiconductors}. The transition from concave to convex Fermi surfaces occurs as a result of the change of sign of the curvature of the sides of the equal-energy contours $v_\perp^{-1}\partial^2 \varepsilon/\partial k_\parallel^2$ (where $v_\perp$ and $k_\parallel$ are the velocity perpendicular and momentum parallel to the contours). We plot $\partial^2 \varepsilon/\partial k_y^2$ for the (a) lowest and (b) second lowest 2D bands along the $\bm{k}\parallel (100)$ direction in Si, Ge, InAs, InSb, GaAs, CdTe and ZnSe quantum wells with infinite square well confinement. The existence of points at which each curve crosses zero indicates the presence of a transition in the Fermi surface geometry which is accompanied by nesting. Note that this transition is absent for the second lowest 2D band in CdTe, indicating that only one nesting density is present, at which the larger Fermi surface is nested.}
\label{sfig:concavity}
\end{figure}

\begin{figure}[h]
\includegraphics[width = 0.995\textwidth]{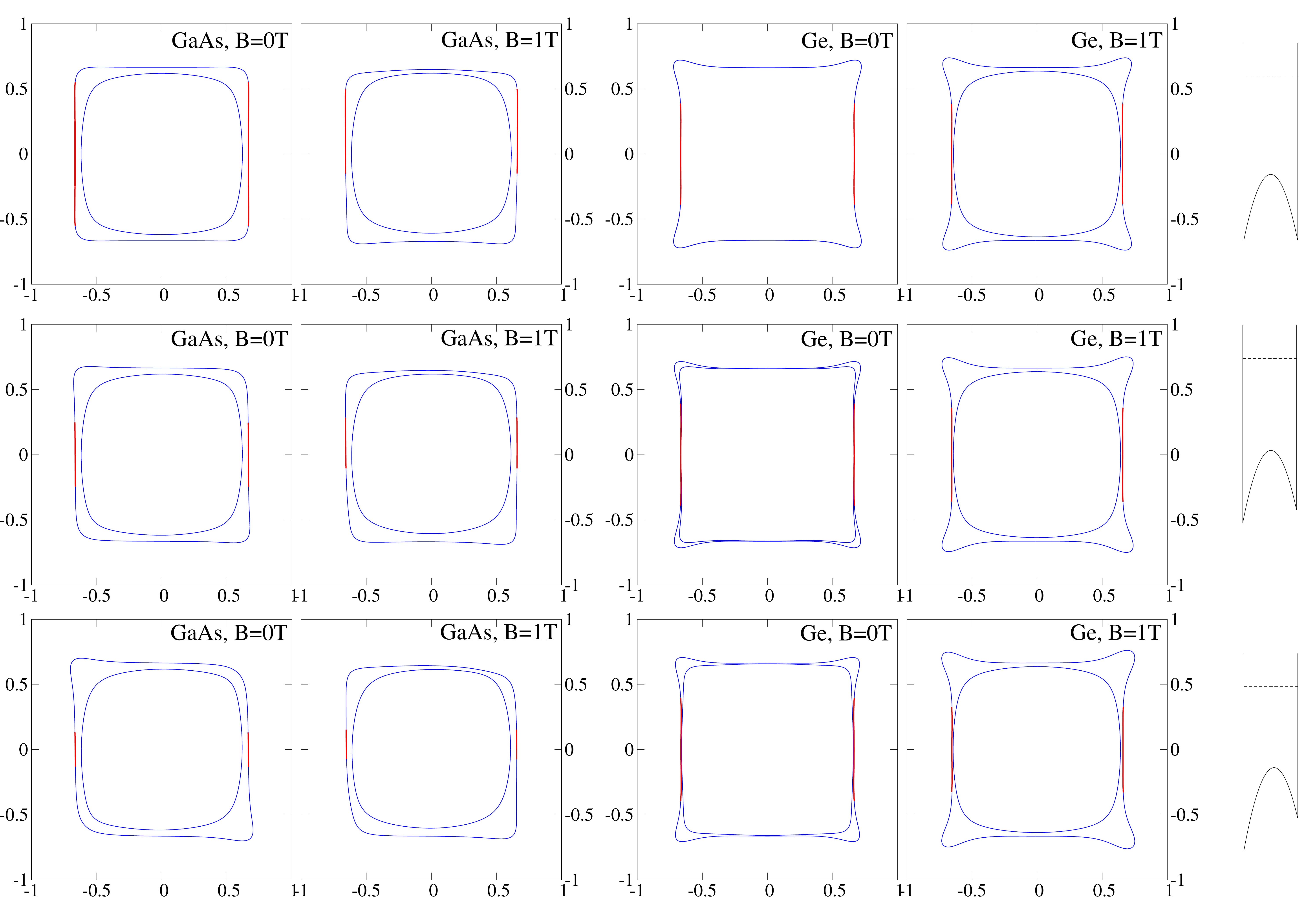} 
\caption{\textbf{Effect of structural inversion asymmetry on nesting in Ge and GaAs wells}. Nested Fermi surfaces for Ge and GaAs quantum wells with increasing asymmetry, either in zero magnetic field or in a 1T magnetic field along (010). Portions of the Fermi surface on which the normal momentum coordinate varies by less than 0.25\% are highlighted in red. The densities chosen are optimal for the nesting of the highlighted pairs of sides. The top row shows Fermi surfaces for a symmetric 20 nm well, the middle row for a 20 nm well with an electric field $E_z = 10^4$ V m$^{-1}$, and the bottom row for a 20 nm well with an electric field $E_z = 2.5\times 10^4$ V m$^{-1}$. The rightmost figures show the confining potential (solid lines) and the transverse quantization energy (dashed lines).}
\label{fig:Ge_wells}
\label{sfig:Rashba}
\end{figure}

\end{document}